\documentclass[12pt,letterpaper]{article}

\usepackage{amsmath,amssymb,calc, latexsym, bbm, array, amsthm}
\usepackage{graphicx}
\usepackage{subfig}
\usepackage[nosort]{cite}
\usepackage{epsfig,psfrag}

\usepackage{ulem}
\usepackage{color}
\definecolor{burgundy}{rgb}{0.8,.2,.2}

\makeatletter
\renewcommand\section{\@startsection {section}{1}{\z@}%
                                   {-3.5ex \@plus -1ex \@minus -.2ex}
                                   {2.3ex \@plus.2ex}%
                                   {\normalfont\large\bfseries}}
\renewcommand\subsection{\@startsection{subsection}{2}{\z@}%
                                     {-3.25ex\@plus -1ex \@minus -.2ex}%
                                     {1.5ex \@plus .2ex}%
                                     {\normalfont\bfseries}}
\makeatother

\theoremstyle{plain}

\newtheorem{lemma}[equation]{Lemma}

\theoremstyle{definition}

\let\non\nonumber

\let\a=\alpha
\let\b=\beta

\let\k=\kappa
\let\l=\lambda

\let\S=\Sigma

\newcommand{\del}{\partial}

\def\one{^{(1)}}

\newcommand{\bea}{\begin{eqnarray}}
\newcommand{\eea}{\end{eqnarray}}
\newcommand{\be}{\begin{equation}}
\newcommand{\ee}{\end{equation}}
\def\ba#1\ea{\begin{align}#1\end{align}}
\def\ban#1\ean{\begin{align*}#1\end{align*}}
\newcommand{\bma}{\begin{pmatrix}}
\newcommand{\ema}{\end{pmatrix}}

\newcommand{\hlf}{\frac{1}{2}}

\newcommand{\F}{{\mathbb F}}

\newcommand{\CC}{{\mathbb C}}

\newcommand{\cF}{{\cal F}}

\newcommand{\G}{\Gamma}

\newcommand{\dd}{\delta}

\newcommand{\f}{\psi}

\newcommand{\D}[1]{\ensuremath{\mathrm{D}#1}}

\newcommand{\C}[1]{$(\ref{#1})$}

\def\IZ{\relax\ifmmode\mathchoice
{\hbox{\cmss Z\kern-.4em Z}}{\hbox{\cmss Z\kern-.4em Z}}
{\lower.9pt\hbox{\cmsss Z\kern-.4em Z}} {\lower1.2pt\hbox{\cmsss
Z\kern-.4em Z}}\else{\cmss Z\kern-.4em Z}\fi}
\def\IR{\relax{\rm I\kern-.18em R}}

\def\one{{\hbox{ 1\kern-.8mm l}}}

\def\tr{{\rm tr\,}}
\def\Tr{{\rm Tr\,}}

\newlength{\bredde}
\def\slash#1{\settowidth{\bredde}{$#1$}\ifmmode\,\raisebox{.15ex}{/}
\hspace*{-\bredde} #1\else$\,\raisebox{.15ex}{/}\hspace*{-\bredde}
#1$\fi}

\newsavebox{\zzzbar}
\sbox{\zzzbar}
  {\setlength{\unitlength}{0.9em}
  \begin{picture}(0.6,0.7)
  \thinlines
  \put(0,0){\line(1,0){0.6}}
  \put(0,0.75){\line(1,0){0.575}}
  \multiput(0,0)(0.0125,0.025){30}{\rule{0.3pt}{0.3pt}}
  \multiput(0.2,0)(0.0125,0.025){30}{\rule{0.3pt}{0.3pt}}
  \put(0,0.75){\line(0,-1){0.15}}
  \put(0.015,0.75){\line(0,-1){0.1}}
  \put(0.03,0.75){\line(0,-1){0.075}}
  \put(0.045,0.75){\line(0,-1){0.05}}
  \put(0.05,0.75){\line(0,-1){0.025}}
  \put(0.6,0){\line(0,1){0.15}}
  \put(0.585,0){\line(0,1){0.1}}
  \put(0.57,0){\line(0,1){0.075}}
  \put(0.555,0){\line(0,1){0.05}}
  \put(0.55,0){\line(0,1){0.025}}
  \end{picture}}

\newcommand{\ena}{\end{eqnarray}}
\newcommand{\beqa}{\begin{eqnarray}}
\newcommand{\eeqa}{\end{eqnarray}}

\def\G{\Gamma}

\renewcommand{\b}{\beta}
\newcommand{\g}{\gamma}

\def\d{{\rm d}}

\newcommand{\ibar}{{\bar \imath}}
\newcommand{\jbar}{{\bar \jmath}}

\newcommand{\kbar}{{\bar k}}

\newfont{\goth}{ygoth.tfm scaled 1200}                   

\def\a{\alpha}
\def\b{\beta}

\def\th{\theta}
\def\f{\phi}
\def\g{\gamma}

\def\j{\psi}
\def\k{\kappa}
\def\l{\lambda}

\def\D{\Delta}
\def\F{\Phi}
\def\G{\Gamma}
\def\J{\Psi}
\def\L{\mathcal{L}}

\def\S{\Sigma}

 \numberwithin{equation}{section}

\def\1{{(1)}}
\def\2{{(2)}}
\def\3{{(3)}}

\def\1{{\bf 1}}
\def\a{{\alpha}}

\def\CC{{\mathbb C}}

\renewcommand\sl{\!\!\!\!\diagup}

\def\cF{{\cal F}}

\begin{document}

\begin{titlepage}
\hbox to \hsize{\hspace*{0 cm}\hbox{\tt }\hss
   \hbox{\small{\tt }}}

\vspace{1 cm}

\center{\bf \Large Conformal Invariance of $(0,2)$ Sigma Models on \\ Calabi-Yau Manifolds}

\vspace{1 cm}

\vspace{1 cm}
\centerline{\large Ian T. Jardine$^\dagger$\footnote{jardinei@physics.utoronto.ca}\, and
 Callum Quigley$^\dagger$\footnote{cquigley@physics.utoronto.ca}}

\vspace{0.5cm}

\centerline{\it ${}^\dagger$Department of Physics, University of Toronto, Toronto, ON M5S 1A7, Canada}

\vspace{0.3 cm}

\begin{abstract}

Long ago, Nemeschansky and Sen demonstrated that the Ricci-flat metric on a Calabi-Yau manifold could be corrected, order by order in perturbation theory, to produce a conformally invariant $(2,2)$ nonlinear sigma model. Here we extend this result to $(0,2)$ sigma models for stable holomorphic vector bundles over Calabi-Yaus.
\end{abstract}

\end{titlepage}

\section{Introduction}

The study of strings propagating on Calabi-Yau backgrounds has a long and rich history. While it was known from early on that Ricci-flat K\"ahler manifolds provided supersymmetric solutions of the classical supergravity equations, there was a period of confusion regarding their fate once $\a'$-corrections were taken into account. On the one hand,~\cite{AlvarezGaume:1985xfa}\ seemed to prove that the Ricci-flat metric of a Calabi-Yau led to a finite nonlinear sigma model (NLSM), with $(2,2)$ supersymmetry, to all orders in worldsheet perturbation theory. However, explicit computation of the $(2,2)$ NLSM beta-function showed that it had a non-zero contribution at four-loop order~\cite{Grisaru:1986px}. The resolution of this seeming paradox was presented in~\cite{Nemeschansky:1986yx}, and boiled down to the freedom of adding finite local counterterms at each order in perturbation theory. In more detail, the authors of~\cite{Nemeschansky:1986yx}\ showed that one could add corrections to the Ricci-flat metric at each loop order, so that the full, all-order beta function is satisfied. This can be interpreted as a (non-local) field redefinition of the metric, or as a modification of the subtraction scheme, so that Ricci-flatness amounts to finiteness to all orders.

The discussion of the previous paragraph was entirely in the context of $(2,2)$ NLSMs on Calabi-Yau backgrounds, but it is very natural to ask how it carries over to the heterotic setting with only $(0,2)$ supersymmetry. In fact, given the intense focus on phenomenological heterotic models at the time of~\cite{Nemeschansky:1986yx}, one easily wonders why such an investigation did not occur decades ago. One likely explanation\footnote{We thank Jacques Distler for explaining some of this historical context and providing his insights.} is that around the same time it was shown that generic $(0,2)$ models suffered from non-perturbative instabilities~\cite{Dine:1986zy}, and so investigating their perturbative renormalization may have seemed inconsequential.\footnote{Investigations early on, however, did not reveal any obstructions at the first few orders~\cite{Witten:1985bz,Witten:1986kg}.} However, we now know that the superpotential contributions from individual worldsheet instantons can actually sum to zero in many $(0,2)$ models~\cite{Distler:1987ee,Basu:2003bq,Beasley:2003fx},\footnote{See, however,~\cite{Bertolini:2014dna}\ for some important limitations on the extent of these claims.} thereby eliminating those instabilities. While this provides renewed motivation for studying the issue of finiteness of $(0,2)$ models on Calabi-Yau backgrounds\footnote{There has also been renewed interest in heterotic geometry from the target space perspective, and the corresponding moduli problem, see e.g.~\cite{Anderson:2014xha,delaOssa:2014cia}.}, during the interim powerful spacetime arguments emerged that answered the question in the affirmative to all orders in perturbation theory~\cite{Dine:1986vd}. In this note, we revisit this question and confirm the expected result directly using worldsheet techniques. In section~\ref{sec:review}\ we review $(0,2)$ models, their beta-functions, and setup our conventions. Section~\ref{sec:proof}\ is the main body of this work where we prove our claim. We summarize our results and outline some open questions in section~\ref{sec:summary}.

\section{Review of the $(0,2)$ NLSM and beta functionals}\label{sec:review}
\subsection{$(0,2)$ superspace and superfields}

We will work in Euclidean $(0,2)$ superspace with coordinates $(z,\bar z, \th, \bar\th)$. We assign $\th$ a $U(1)_R$ charge $q_R=+1$ and $\bar\th$ a charge $q_R=-1$, and take the fermionic integration measure such that $\int d^2\th\, \th\bar\th =1$. The covariant derivatives and supercharges are given by:
\ba\label{eqn:DandQ}
&D =\del_\th +\bar\th\bar\del,\qquad \bar D = \del_{\bar\th} +\th \bar\del,\\
&Q= -\del_\th +\bar\th\bar\del,\qquad \bar Q = -\del_{\bar\th} +\th \bar\del,
\ea
with non-trivial anti-commutators
\ba\label{eqn:algebra}
\{D,\bar D\} = 2\bar\del,\qquad \{Q,\bar Q\} = -2\bar\del.
\ea
Chiral fields are annihilated by $\bar D$, and the scalar and Fermi chiral fields have the component expansions
\ba\label{eqn:chiral}
\F = \f+\sqrt{2}\th\psi +\th\bar\th \bar\del \f,\quad \G =\g +\sqrt{2}\th F +\th\bar\th \bar\del\g.
\ea
Anti-chiral fields on the other hand are annihilated by $D$, and they have corresponding component expansions
\ba\label{eqn:anti-chiral}
\bar\F = \bar\f-\sqrt{2}\bar\th\bar\psi -\th\bar\th \bar\del \f,\quad \bar\G =\bar\g +\sqrt{2}\bar\th \bar F-\th\bar\th \bar\del\bar\g.
\ea
We assign $q_R=0$ to all the (lowest components of the) chiral fields. To simplify the details, we will demand the existence of a non-anomalous global $U(1)_L$ symmetry, under which $\F,\bar\F,\G,\bar\G$ have charges $q_L=0,0,+1,-1$, respectively. It is straightforward to generalize to cases without the $U(1)_L$ symmetry.

\subsection{The classical $(0,2)$ NLSM}\label{sub:NLSM}
The most general, renormalizable, $U(1)_R\times U(1)_L$ invariant, $(0,2)$ supersymmetric action is given by
\ba\label{eqn:02NLSM}
S_{NLSM} = -\frac{1}{4\pi\a'}\int d^2z d^2\th\left[\tfrac12\left(K_i(\F,\bar\F)\del\F^i -K_\ibar(\F,\bar\F)\del \bar\F^\ibar \right) -h_{a\bar b}(\F,\bar\F)\bar\G^{\bar b}\G^ a\right].
\ea
The $(1,0)$-form $K = K_i\d\f^i$ (with complex conjugate  $K^*=K_\ibar\d\f^\ibar$) is the $(0,2)$ analog of the $(2,2)$ K\"ahler potential, and is only defined up to shifts by holomorphic $(1,0)$-forms: $K(\F,\bar{\F}) \rightarrow K(\F,\bar{\F}) + K'(\F)$. The action is also invariant under $K \rightarrow K +i\,\del f$, for any real-valued function $f$. Holomorphic redefinitions of the Fermi fields, $\G^a \rightarrow M^a{}_b(\F)\G^b$, imply that the Hermitian metric $h_{a\bar b}$ is only defined up to transformations of the form $h\rightarrow M^\dag h M$.\footnote{Non-Hermitian metrics, with components $h_{ab}$ and $h_{\bar a\bar b}$, are forbidden by the $U(1)_L$ symmetry.} In particular, rescaling $h_{a\bar b}$ by a constant factor leads to an equivalent theory; this will be relevant at a later stage.

Expanding the supersymmetric sigma model action~\C{eqn:02NLSM}\ in components,\footnote{The auxiliary fields appear in the action $\L_{aux} = (F^a + A^a_{i\ b}\psi^i\g^a)h_{a\bar b}(\bar{F}^{\bar b} - A^{\bar b}_{\jbar\ \bar{e}}\bar\psi^\jbar\g^{\bar{e}}),$ which vanishes on shell.} we obtain
\ba\label{eqn:component}
S_{NLSM} =\frac{1}{2\pi\a'}\int d^2z&\left[ \tfrac12 g_{i\jbar}\left(\del\f^i\bar\del\bar\f^\jbar  + \bar\del\f^i\del\bar\f^\jbar\right)+ \tfrac12 B_{i\jbar}\left(\del\f^i\bar\del\bar\f^\jbar  - \bar\del\f^i\del\bar\f^\jbar\right)\right.\non\\
&+ g_{i\jbar}\bar\psi^\jbar\left(\del\j^i +\left(\G^{-i}_{j k}\del\f^j +\G^{-i}_{\jbar k} \del\f^{\jbar}\right)\psi^k\right) \non\\
&\left.+ h_{a\bar b}\bar\g^{\bar b} \left(\bar\del\g^a + A^a_{i\,b} \bar\del\f^i\g^b\right) +  \cF_{i\jbar a\bar b}\j^i\bar\psi^\jbar  \bar\g^{\bar b} \g^a \right],
\ea
where the couplings in the above action are determined by $K_i$ and $h_{a\bar b}$:
\ba\label{eqn:couplings}
g_{i\jbar} = \del_{(i} K_{\jbar)}, \qquad  &B_{i\jbar} = \del_{[\jbar} K_{i]},\qquad A^a_{i\, b} = h^{a\bar{a}}\del_i h_{b\bar{a}}, \qquad \cF_{i\jbar a\bar b} = h_{a \bar a}\del_i A^{\bar a}_{\jbar\ \bar b}\non,
\ea
and the connection $\G^-$ is defined by $\G^{\pm} = \G \pm\tfrac12 H,$
where $\G$ is the usual Christoffel connection for the metric $g$ and $H=dB$ is the tree-level torsion. In particular, we have
\ba
\G^{-i}_{jk} = \G^i_{jk} -\tfrac12 H^i{}_{jk} = g^{i\jbar}\del_j g_{k\jbar},\qquad \G^{-i}_{\jbar k} =  \G^i_{\jbar k} -\tfrac12 H^i{}_{\jbar k} = -H^i{}_{\jbar k},
\ea
along with their complex conjugates, and the rest of the components vanish.

Geometrically, the sigma model action describes maps of a two-dimensional worldsheet $\S$ into a target manifold $M$, equipped with a metric, $g$, and $B$-field.
$(0,2)$ supersymmetry guarantees that $M$ is a complex manifold~\cite{Hull:1985jv,Sen:1986mg}, with fundamental form $\omega = ig_{i\jbar}\,d\f^i d\bar\f^\jbar,$ related to $H$ by
\be\label{eqn:susyreln}
H = dB = i(\bar\del- \del)\omega.
\ee
The right-moving fermions, $\psi$, transform as sections of (the pullback of) $TM$, coupled to the $H$-twisted connection $\G^-$. The left-moving fermions, $\g$, transform as sections of a (stable) holomorphic bundle $E\rightarrow M$, where $E$ is equipped with a Hermitian metric $h_{\a\bar\b}$, and holomorphic connection $A_i$ with curvature ${\cal F}_{i\jbar}$. To simplify our analysis we will assume that $E$ is stable, as opposed to just poly-stable or semi-stable. The fact that $U(1)_L$ is anomaly-free translates into the statement that $c_1(E)=0$. Similarly, if $U(1)_R$ is anomaly-free then $c_1(M)=0$ and the theory flows to a non-trivial $(0,2)$ SCFT in the infrared.

\subsection{Beta functionals of the (0,2) NLSM}
The one-loop beta functionals for $(0,2)$ NLSMs were derived in~\cite{Nibbelink:2012wb}\ by demanding $(0,2)$ super-Weyl invariance of the quantum effective action. Crucially, they found three beta functions for the theory, which are in 1-1 correspondence with the spacetime BPS equations of the heterotic string:
\ba
\dd\J_M &= \nabla_M^{(-)}\epsilon = 0, \label{eqn:gravitino}\\
\dd \l &= \left(\del\sl\varphi-\tfrac12 H\!\sl\right)\epsilon=0, \label{eqn:dilatino}\\
\dd\chi &= \cF\!\!\sl\epsilon=0 \label{eqn:gaugino}.
\ea
Given the relation between $(0,2)$ worldsheet and $N=1$ spacetime supersymmetries, it is rather natural that manifestly $(0,2)$ supersymmetric beta functions are related to the spacetime BPS equations.

We will restrict ourselves to flat worldsheets, $\S=\CC$, and therefore will not be sensitive to the coupling of the dilaton or its corresponding beta functional. In other words, we will only study the conditions for finiteness of the SCFTs that the NLSMs flow to, rather than tackle the more involved question of (super-)Weyl invariance of the full heterotic string worldsheet. Therefore, the $(0,2)$ NLSMs we consider are determined by two only coupling functionals: $K_i$ and $h_{a\bar b}$, each with their corresponding beta functional $\b^K_i$ and $\b^h_{a\bar b}$. Let us write $\b^{(1)}$ for the one-loop contribution to the beta functionals and $\D\b$ for the sum of all higher loop contributions: $\b = \b^{(1)} + \D\b$.\footnote{While the $\D\b$ are scheme dependent, once we fix a renormalization scheme their expressions are unique.} We will compare our approach to~\cite{Nibbelink:2012wb}\ further in what follows. For our starting point, we will take the one-loop NLSM beta functionals to be given by
\ba
\b_i^{K(1)} &= c\, g^{j\kbar}(\del_j g_{i\kbar} - H_{ij\kbar})  \label{eqn:beta K}\\
\b_{a\bar b}^{h(1)} &= c'\, g^{i\jbar} \cF_{i\jbar a\bar b} \label{eqn:beta h},
\ea
where $c$ and $c'$ are some known constants whose precise values we do not require. Our $\b^{(1)}_h$ coincides with that of~\cite{Nibbelink:2012wb}, and its vanishing clearly implies the spacetime gaugino equation~\C{eqn:gaugino}. Our $\b^{(1)}_K$ is a little more subtle, because it corresponds to a linear combination of the two remaining beta functionals of~\cite{Nibbelink:2012wb}.\footnote{It bears pointing out that $i\b^{K(1)}_i = i(\G^{-j}_{ij} - \G^{-\jbar}_{i\jbar}):=\G^-_i$ is the induced connection on the canonical bundle~\cite{Gillard:2003jh}. When $\G^-_i$ is flat, then $\nabla^{(-)}$ has $SU(n)$ holonomy as required by the gravitino equation~\C{eqn:gravitino}.} One member of that pair reads
\ba\label{eqn:dilatino2}
\del_i\varphi = \tfrac12 H_{ij\bar k}g^{j\bar k},
\ea
where $\varphi$ is the dilaton field, and this is equivalent to the dilatino equation~\C{eqn:dilatino}. Since our interest is in $(0,2)$ SCFTs defined on the plane, it should come as no surprise that we are not sensitive to~\C{eqn:dilatino2}. Instead, we will treat~\C{eqn:dilatino2}\ as a constraint that {\it defines} $\varphi$ for our models (at least at leading order in $\a'$). We remark that~\C{eqn:dilatino2}\ can always be solved locally, and so it is indeed a valid definition of $\varphi$ in a CFT. If we wish to promote such a solution to string theory, we must further ensure that $\varphi$ is globally defined, and this may obstruct such a lift.\footnote{For example, this occurs for torsional NLSMs on $S^3\times S^1$. See Appendix C of~\cite{Quigley:2012gq}\ for more details.} Fortunately, when including worldsheet supergravity the field $\varphi$ is intrinsically defined by its coupling to the worldsheet curvature, and this issue is avoided.

Using the constraint~\C{eqn:dilatino2}, it is straightforward to show that our choice of $\b^{(1)}_K$ reproduces the correct beta functionals for the physical couplings $g$ and $B$:
\ba
\b^{g(1)}_{i\jbar} &= R_{i\jbar} +2\nabla_i\nabla_\jbar \varphi -\frac14 H_{iMN}H_\jbar{}^{MN} = -\del_{(i}\b^{K(1)}_{\jbar)},\\
\b^{B(1)}_{i\jbar} &= \nabla_M\left(e^{-2\varphi}H^M{}_{i\jbar}\right) = \del_{[i}\b^{g(1)}_{\jbar]}.
\ea
We take this result as support of our starting point~\C{eqn:beta K}-\C{eqn:beta h}.

We now wish to state the $(0,2)$ generalization of the key lemma from~\cite{Nemeschansky:1986yx}.
\begin{lemma}
  Let $E\rightarrow M$ be a stable holomorphic vector bundle with data $(\tilde g, \tilde B, \tilde A)$, derived from $\tilde K$ and $\tilde h$, with vanishing one-loop beta functionals. Then, there exist couplings
  \ba
  K_i = \tilde K_i -\dd K_i, \qquad   h_{a\bar b} = \tilde h_{a\bar b} -\dd h_{a\bar b} \non,
  \ea
  with corresponding data $(g,B,A)$  such that
  \ba
  \b^K(g,B,A) &= \b^{K(1)}(g,B) + \D\b^K(g,B,A) = \b^{K(1)}(\tilde g, \tilde B), \label{eqn:lemma1}\\
  \b^h(g,B,A) &= \b^{h(1)}(g,A) + \D\b^h(g,B,A) = \b^{h(1)}(\tilde g, \tilde A),\label{eqn:lemma2}
  \ea
  In particular, the full beta-functionals for $(g,B,A)$ can be made to vanish.
\end{lemma}

\section{Proving the lemma}\label{sec:proof}
\subsection{Restriction on $H$}
The validity of a perturbative loop/$\a'$ expansion for a general $(0,2)$ NLSM is doubtful at best. It has been shown by many authors that the existence of non-vanishing $H$-flux in the tree-level action leads to string scale cycles in the geometry, and therefore a breakdown of a large-volume expansion~\cite{strominger-torsion,Becker:2002sx,Becker:2003sh,Melnikov:2014ywa,delaOssa:2014msa}.
Therefore, we will content ourselves with the more conservative goal of proving the above lemma in cases where $H$ vanishes classically. More precisely, we will assume the validity of $\a'$ perturbation theory, where the $\a'\rightarrow0$ limit is smooth\footnote{This rules out the only known class of truly torsional backgrounds, based on the total space $T^2\rightarrow K3$, discovered in~\cite{Dasgupta:1999ss}, along with their related generalizations.}  and that $H\rightarrow0$ in this limit. In such situations, it well known that $M$ is Calabi-Yau to leading order in $\a'$. However, except in the case of the standard embedding with $A=\G$, $H$-flux will be generated by loop effects such as the Bianchi identity,
\ba
dH = \frac{\a'}{4}(\tr R^+\wedge R^+ - \tr\cF\wedge\cF),
\ea
and~\C{eqn:susyreln} will ensure that the full metric, $g$, is no longer K\"ahler.

\subsection{Proving the lemma locally}
In the simplified setting where $H=0$ classically,
it is straightforward to prove the lemma. The general approach will be within the framework of $\a'$ perturbation theory. We will first show that the lemma holds to lowest order in $\a'$, and then extend this to higher orders inductively by using the results of the earlier steps. In this way, we will prove the lemma to all order in $\a'$.

We will first tackle $\b^K$. First, note that~\C{eqn:susyreln}\ allows us to write
\ba
\b^{K(1)}_i = c\left(\del_i\log\det g - 2H_{ij\bar k}g^{j\bar k}\right) = c\left(\del_i\log\det g - 4\del_{[i} g_{j]\bar k} g^{j\bar k}\right).
\ea
Next, we can rewrite~\C{eqn:lemma1}\ as
\ba
c^{-1} \D\b^K_i &= \del_i \log\det(\tilde g/g) + 4 g^{j\bar k}\del_{[i}g_{j]\bar k} \non \\
&= -\del_i \Tr \log\left(\mathbf{1}-\tilde g^{-1}\dd g\right) -4 g^{j\bar k}\del_{[i}\dd g_{j]\bar k} \label{eqn:deltabeta}\\
& = \sum_{n>0}\frac{1}{n}\, \del_i\, \Tr \! \left(\tilde g^{-1}\dd g\right)^n -4 \sum_{n\geq0} \tilde g^{j\bar\ell} \left[\left(\tilde g^{-1}\dd g\right)^n\right]^{\bar k}_{\bar\ell} \del_{[i}\dd g_{j]\bar k} , \non
\ea
where we have expanded $g = \tilde g -\dd g$ and used the fact that $\tilde g$ is K\"ahler. As in~\cite{Nemeschansky:1986yx}, this equation may be solved iteratively for $\dd K$, order by order in $\a'$, in terms of the input data $c^{-1}\D\b^K$.\footnote{For instance, the second-order correction will depend on the tree-level data and quadratically on the one-loop correction.} To lowest order, this equation is simply
\ba
c^{-1} \D\b^K_i &= \del_i(\tilde g^{j\bar k} \dd g_{j\bar k}) -4\tilde g^{j\bar k} \del_{[i}\dd g_{j]\bar k} = \del_i \left(\tilde g^{j\bar k} \del_{(j}\dd K_{\bar k)}\right)+ 2 \tilde g^{j\bar k}\del_{\bar k} \dd B_{ij},
\ea
where we defined $\dd B_{ij} = \del_{[j}\dd K_{i]}$. We can see that the righthand side is the sum of $\del$-exact and co-exact pieces:
\ba
c^{-1}\D\b^K = \tfrac12\,\del\left(\tilde\nabla\cdot\dd K\right) + 2 \del^\dag B,\label{eqn:decomp1}
\ea
where $\del^\dag = *\bar\del*$ is the adjoint of $\del$. Because the target space $M$ is simply connected, the Hodge decomposition theorem tells us that every 1-form has such a decomposition. In particular, the left hand side can be expressed as
\ba
c^{-1} \D\b^K = \del f + 2\del^\dag \chi \label{eqn:decomp2}
\ea
for some scalar function $f$ and a $(2,0)$-form $\chi$. The factor of 2 is merely used for convenience. Note that $f$ is only defined up to the addition of a constant, while $\chi$ is only well-defined modulo a co-closed 2-form, $\gamma$. Applying the Hodge decomposition once again to $\gamma$, we see that $\gamma$ must be co-exact, and so $\chi\sim\chi +\del^\dag\rho$ for $\rho$ a $(3,0)$-form.

Clearly, the equations~\C{eqn:decomp1}\ and~\C{eqn:decomp2}\ are equivalent to the pair
\ba
& \tfrac12\tilde\nabla\cdot\dd K = f, \qquad \del (\dd K) = \chi,\label{eqn:div-curl}
\ea
where again $f\sim f+const.$ and $\chi\sim\chi+\del^\dag\rho$. On a simply connected manifold, any 1-form is completely determined by its divergence and exterior derivative, therefore $\dd K$ is in principle fixed to lowest order. To see this in detail, first we can decompose $\dd K$ itself as
\ba
\dd K = \del k + \del^\dag\kappa.
\ea
We can assume $k$ is real scalar function, since any imaginary component only adds a total derivative to the action, as noted below~\C{eqn:02NLSM}. Then we have
\ba
\tfrac12\tilde\nabla\cdot\dd K = \tfrac12\left(\del^\dag\dd K +\bar\del^\dag\dd\bar K\right) = \tfrac12\left(\del^\dag \del +\bar\del^\dag\bar\del\right)k = \tfrac12\left(\D_\del +\D_{\bar\del}\right)k =\D_\del k,
\ea
because $\tilde g$ is K\"ahler and on a K\"ahler manifold $\D_\del = \D_{\bar\del} = \hlf\D_d$. Here we have introduced the Laplace-Beltrami operator $\D_\del = \del\del^\dag + \del^\dag\del$, with similar expression for the other Laplacians.
Next, we can use the ambiguity $\k\sim \k+ \del^\dag\a$ to remove the co-exact piece of $\k$ from its Hodge decomposition, ensuring that $\k$ is $\del$-exact and therefore closed. In other words, $\del(\dd K) = \del\del^\dag\k = \D_\del\k$, and so~\C{eqn:div-curl}\ can be written
\ba
\D_\del k = f,\qquad \D_\del\k = \chi.
\ea
These equations can always be solved locally by inverting the Laplacian, provided $f$ and $\chi$ contain no zero-modes of $\D_\del$. For $\chi$ this is trivial, as there are no harmonic $(2,0)$ forms on $M$, while for the function $f$ we can use the ambiguity $f\sim f+const$ to ensure that $f$ has no constant term. We will discuss the existence of global solutions in the following section. Thus we obtain the lowest order to solution for $\dd K$:
\ba
\dd K = \del(\D^{-1}_\del f) + \del^\dag(\D^{-1}_\del \chi),
\ea
where $f$ and $\chi$ are determined from the input data $c^{-1}\D\b^K$ via~\C{eqn:decomp2}.
Plugging this back into~\C{eqn:deltabeta}, we can iteratively solve for $\dd K$ order by order in the expansion.

Turning now to $\beta^h$, in a similar fashion we rewrite~\C{eqn:lemma2}\ as
\ba
c'^{-1}\D\b^h_{a\bar b} &= \tilde g^{i\jbar} \tilde{\cal F}_{i\jbar a \bar b} - g^{i\jbar} {\cal F}_{i\jbar a \bar b} \label{eqn:deltabeta_h}\\
&= \tilde g^{i\jbar} \tilde{\cal F}_{i\jbar a \bar b} -\sum_{m,n\geq0} \tilde g^{i\bar k}\left[\left(\tilde g^{-1}\dd g \right)^m\right]^\jbar_{\bar k}(\tilde h_{a\bar a}-\dd h_{a\bar a})\del_i\left(\tilde h^{\bar a c}\left[\left(\tilde h^{-1}\dd h\right)^n\right]^b_c\del_\jbar(\tilde h_{b\bar b}-\dd h_{b\bar b})\right).\non
\ea
To lowest orders in $\dd h$ and $\dd g$, this can be written as
\ba\label{eqn:delta_h}
D_A^2 \dd h_{a\bar b} = c'^{-1}\D \b^h_{a\bar b} + \tilde{\cal F}^{i\jbar}{}_{a\bar b}\dd g_{i\jbar},
\ea
where $D_A^2$ is gauge-covariant Laplacian, and we have used $\b^{h(1)}_{a\bar b}(\tilde g,\tilde A) =\tilde g^{i\jbar}{\tilde\cal F}_{i\jbar a\bar b} =0$. Since we have assumed the bundle $E$ is stable, we have $\dim H^0(End~E)=1$ (see, for example, Cor.~1.2.8 of~\cite{huybrechts_lehn_2010}), so $D_A$ has a unique zero-mode corresponding to the uniform rescaling of $h_{a\bar b}$. However, as mentioned in section~\ref{sub:NLSM}, such scaling of the bundle metric can be absorbed into the Fermi fields. So up to this unphysical ambiguity, we can solve~\C{eqn:delta_h}\ (locally) to find $\dd h$ uniquely in terms of the lowest order solution for $\dd g$ from the previous paragraphs and the input $\D\b^h$. As before, the lowest order solution can be plugged back into~\C{eqn:deltabeta_h}\ to get an iterative solution for $\dd h$.

In summary, we have demonstrated how to construct local, perturbative solutions for $\dd K $ and $\dd h$, order by order in the $\a'$/loop expansion, in terms of $\D\b^K$ and $\D\b^h$. In the following section we will show that these local solutions actually patch together into well-defined global ones. This will complete the proof that the corrected sigma model data, $K=\tilde K -\dd K$ and $h = \tilde h -\dd h$, have vanishing beta functionals.

\subsection{Proving the lemma globally}

With the existence of a local solution to the set of equations defining $K_i$ and $h_{a\bar{b}}$ shown, we will now argue that this solution is globally defined. The key point to notice is that the solutions for $\dd K_i$ and $\dd h_{a\bar b}$ are determined entirely in terms of the higher loop beta functionals $\D\b^K$ and $\D\b^h$. We will now show that the $\D\b$ are globally defined, so that $\dd K$ and $\dd h$ are as well. The arguments presented in this section closely parallel original ones of~\cite{Nemeschansky:1986yx}\ for the simpler $(2,2)$ case.

We will begin with $\D\b^K$. First, recall that
\begin{equation}
\beta^g_{i\bar{\jmath}}=-\partial_{(\bar{\jmath}}\beta_{i)}^K
\end{equation}
must have a covariant form, since it provides the metric equations of motion, and so defines a global tensor on $M$. On the other hand, $\b^K$ itself need not be globally defined, as a shift by a holomorphic one-form is allowed. So when we go from one patch to another, $\Delta\beta^K$ must satisfy:
\begin{align}
\Delta\beta^K_i{}'-\frac{\partial z^j}{\partial z'^i}\Delta\beta^K_j=f_i(z),\label{eqn:patchingK}\\
\Delta\beta^K_{\bar{\jmath}}{}'-\frac{\partial z^{\bar{\imath}}}{\partial z'^{\bar{\jmath}}}\Delta\beta^K_{\bar{i}}=g_{\bar{\jmath}}(\bar{z}),
\end{align}
for set of (anti-)holomorphic functions $f_i$ and $g_\jbar$.
Furthermore, we must consider the possibility for gauge transformation between the patches. Since $\beta^g_{i\bar{\jmath}}$ is a globally defined singlet, then $\partial_{(\bar{\jmath}}\Delta\beta_{i)}$ must also be a globally defined singlet. So we get the relationship between the two patches from a gauge transformation as
\begin{align}
\Delta\beta^K_i{}'-\Delta\beta^K_i= \tilde f_i(z),\label{eqn:patchingK2}\\
\Delta\beta^K_{\bar{\jmath}}{}'-\Delta\beta^K_{\bar{j}}= \tilde g_{\bar{\jmath}}(\bar{z}),
\end{align}
for some other (anti-)holomorphic functions $\tilde f$ and $\tilde g$.

Let us focus on $\D\b^K_i$, though similar arguments apply for $\D\b^K_{\bar{\jmath}}$. On general grounds, we expect $\D\b^K$ to be expressible as a product of derivatives of $K_i$, $K_\jbar$ and $h_{a\bar b}$, with indices contracted by the metrics $g^{i\jbar}$ and $h^{a\bar b}$. There are four classes of factors that could make up our beta function,
\begin{align}\label{term1}
\partial_{k_1}...\partial_{k_n}\partial_{\bar{l}_1}...\partial_{\bar{l}_m}K_i, \\ \label{term2}
\partial_{i}\partial_{k_1}...\partial_{k_n}\partial_{\bar{l}_1}...\partial_{\bar{l}_m}K_{k_{n+1}}, \\ \label{term3}
\partial_{i}\partial_{k_1}...\partial_{k_n}\partial_{\bar{l}_1}...\partial_{\bar{l}_m}K_{\bar{l}_{m+1}}, \\ \label{term4}
\partial_{i}\partial_{k_1}...\partial_{k_n}\partial_{\bar{l}_1}...\partial_{\bar{l}_m}h_{a\bar{b}}.
\end{align}
To simplify our analysis, we will make one further reasonable assumption:
the $B$-field, $B_{i\jbar} = \del_{[\jbar}K_{i]}$, can only appear in the beta functions through the invariant combination $H = dB +\tfrac{\a'}{4}(CS(\omega^+)-CS(A))+O(\a'^2)$. So that even if individual factors in the list above have anomalous transformation properties, because of the Green-Schwarz mechanism, these effects will cancel out of the full beta functions. Thus we will not need to examine anomalous transformations of the above factors, and we can restrict our attention to conventional diffeomorphisms and gauge transformations only.

When we consider a spacetime transformation between patches of one of the factors listed above, two types of terms can arise. We will follow~\cite{Nemeschansky:1986yx}\ and refer to these types as homogeneous and inhomogeneous. Homogeneous terms are the standard tensor transformations, with each spacetime index contracted with a Jacobian matrix, while the inhomogeneous terms involve higher derivatives between the coordinates on the patches. Since we contract all the indices except $i$ using $g^{j\bar k}$, the homogeneous terms will all cancel, save for a single Jacobian factor associated with the uncontracted $i$ index. 
In other words, the homogenous terms satisfy the transformation law~\C{eqn:patchingK}\ with $f_i=0$.

What about the inhomogeneous terms, can they generate a non-zero $f_i$? Note that an inhomogeneous term with a factor of the form
\begin{equation}
\frac{\partial^rz'^k}{\partial z^{j_1}...\partial z^{j_r}}
\end{equation}
will have more indices on the bottom then on the top. Since we must contract all indices except $i$, this implies that the inhomogeneous terms must have factors involving the inverse metric. This is a function of both $z$ and $\bar{z}$, which cannot appear on the righthand side of~\C{eqn:patchingK}. However, we could consider inhomogenous terms with a factor of the form
\begin{equation}
\frac{\partial^rz'^k}{\partial z^{i}\partial z^{j_1}...\partial z^{j_{r-1}}}\label{badinhom}.
\end{equation}
If $r>1$, then the same argument from above applies. However, we could have a term that involves
\begin{equation}
\frac{\partial^2z'^k}{\partial z^{i}\partial z^{j}}
\end{equation}
Since this has a balanced set of indices, we can imagine contracting this with a function with only holomorphic dependence. The only term available would be another Jacobian factor, giving us two possibilities. The first would involve a term like
\begin{equation}
\frac{\partial^2z'^k}{\partial z^{i}\partial z^{j}}\frac{\partial z^j}{\partial z'^{k}}=\frac{\partial^2z'^k}{\partial z^{i}\partial z'^{k}}=0.
\end{equation}
The other would be
\begin{equation}
\frac{\partial^2z'^k}{\partial z^{i}\partial z^{j}}\frac{\partial z'^j}{\partial z^{k}}.
\end{equation}
However, this type of term would not show up from a coordinate transformation, as we must have $z'$ contracting with $z'$ and $z$ contracting with $z$. So, the inhomogeneous terms involving~\C{badinhom}\ do not show up. Since the homogeneous term implies $f_i(z)=0$ and the inhomogeneous terms cannot contribute a non-zero $f_i$, then we conclude that $\Delta\beta^K$ is globally defined under diffeomorphisms.

We now turn to gauge transformations. Note that terms involving~\C{term1}-\C{term3}\ do not involve gauge indices and so will not transform under gauge transformations. This leaves only terms involving~\C{term4}. Note that $\Delta\beta^K$ does not have any gauge indices so we must contract up terms with~\C{term4}\ with factors of $h^{a\bar{b}}$. Under a gauge transformation, we recall that
\begin{equation}
h_{a\bar{b}}'=U_{a}^{c}h_{c\bar{d}}U_{\bar{b}}^{\bar{d}}.
\end{equation}
So now we consider the full term
\begin{equation}
\partial_{i}\partial_{k_1}...\partial_{k_n}\partial_{\bar{l}_1}...\partial_{\bar{l}_m}(U_{a}^{c}h_{c\bar{d}}U_{\bar{b}}^{\bar{d}}).
\end{equation}
We note that undifferentiated factors of $U$ will cancel against the transformation of $h^{a\bar b}$, and the difference between patches of such terms will vanish. So we only have to worry about terms that involve derivatives acting on $U$. Note, however, that the all derivatives except $\del_i$ must be contracted with $g^{j\bar k}$, which we have noted before is not holomorphic and would violate~\C{eqn:patchingK2}. A factor of the form $\del_i h_{a\bar b}$, however, are not ruled out. Therefore, the only terms that could possibly lead to non-zero holomorphic differences are contractions of $h^{a\bar{b}},h_{a\bar{b}},$ and $A^a_{ib}$. The only terms one can construct from these basic building blocks with the correct index structure are, schematically, $\del_i\tr(hh...h)=0$ and $\tr(A_i h...h)=0$, assuming that gauge group is semi-simple. So we have shown that $\Delta\beta^K$ is a globally defined gauge singlet, as well.

We now go through the same arguments for $\Delta\beta^h$. Analogous to the previous case, we may insist that
\begin{align}
\beta^A_{ia\bar{b}}=\partial_{i}\beta_{a\bar{b}}^h\qquad {\rm and} \qquad \beta^A_{\bar{\jmath}a\bar{b}}=\partial_{\bar{\jmath}}\beta_{a\bar b}^h
\end{align}
are globally defined. Satisfying both of these constraints restricts the patching condition for $\D\b^h$ to be:
\begin{equation}
\Delta\beta^h_{a\bar{b}}{}'-\Delta\beta_{a\bar{b}}=C_{a\bar b}
\end{equation}
where $C$ is a constant matrix. Now we have three classes of factors that could appear in our beta function:
\begin{align}\label{term1h}
\partial_{k_1}...\partial_{k_n}\partial_{\bar{l}_1}...\partial_{\bar{l}_m}K_{k_{n+1}}, \\ \label{term2h}
\partial_{k_1}...\partial_{k_n}\partial_{\bar{l}_1}...\partial_{\bar{l}_m}K_{\bar{l}_{m+1}}, \\ \label{term3h}
\partial_{k_1}...\partial_{k_n}\partial_{\bar{l}_1}...\partial_{\bar{l}_m}h_{a\bar{b}}.
\end{align}
The same argument as before applies for diffeomorphisms. In fact it is even simpler, as all of the spacetime indices must be contracted, and the argument is very similar to the original one found in~\cite{Nemeschansky:1986yx}.
Gauge transformations will be a bit different. Since the factors~\C{term1h}\ and~\C{term2h}\ do not have any gauge dependence, they are invariant under gauge transformations. So we will only need to worry about~\C{term3h}. However, we can see relatively immediately that there are no constant matrices $C_{a\bar b}$ that transform under the correct representation, except the trivial case $C_{a\bar b}=0$.

In summary, we have shown that $\Delta\beta^K$, and $\Delta\beta^h$ are globally defined tensors, under both spacetime diffeomorphisms and gauge transformations. Our proof required assuming that the full beta functions $\b^K$ and $\b^h$ only depend on the $B$-field though the invariant three-form $H$. We do not expect this assumption to affect our result.
Since the $\D\b$ are the data that determine $\dd K$ and $\dd h$, we conclude that these corrections are also globally defined. Thus, the full beta functions for the corrected couplings $K = \tilde K - \dd K$ and $h = \tilde h - \dd h$ can be made to vanish.

\section{Summary and outlook}\label{sec:summary}

In this short note, we have shown that the $(0,2)$ NLSMs on Calabi-Yau backgrounds, equipped with a stable holomorphic bundle, can be made to be finite to all orders in perturbation theory. We have closely paralleled the approach of~\cite{Nemeschansky:1986yx}\ in the $(2,2)$ setting, by demonstrating the existence of a set of counterterms to the sigma model couplings that make the beta functions vanish to all orders.

There are a few unresolved issues, that we leave as open problems for future study. The first question is how our results carry over when the $(0,2)$ NLSMs are coupled to worldsheet supergravity. This will require studying the full Weyl symmetry, along with couplings to the dilaton field and its associated beta function. This would appear to be a prerequisite for going beyond the restriction of demanding $H=0$ at the classical level, and only allowing for $H$ to be generated by loop effects. While tree-level $H$-flux is problematic for a controlled $\a'$-expansion, it may be possible to avoid these complications since we are only interested in the general structure of the theory, and not necessarily perturbing about some particular torsional background. Finally, we have assumed that anomalous transformations will not appear in the beta functions. While we believe this to be true, a cautious reader might be wary. Since the anomaly only shows up at one-loop, they could potentially affect the two-loop beta functions and this should be checked directly.

\section*{Acknowledgements}

This project originated from a discussion between Jacques Distler, Ilarion Melnikov and CQ at the workshop \textit{Heterotic Strings and (0,2) QFT}, hosted by the Mitchell Institute at TAMU in April, 2014. CQ would like the thank the organizers and participants for providing a stimulating and exciting environment. We especially thank Ilarion Melnikov for helpful suggestions at various stages of this paper. We also thank Marc-Antoine Fiset, Stefan Groot Nibbelink, and Eirik Svanes for helpful discussions and feedback on earlier drafts. IJ is supported by a NSERC Discovery grant. CQ was supported by a NSERC fellowship.


\end{document}